# A portable magnetic field of > 3 T generated by the flux jump assisted, pulsed field magnetisation of bulk superconductors

Difan Zhou[1], Mark D. Ainslie[1], Yunhua Shi[1], Anthony R. Dennis[1], Kaiyuan Huang[1], John R. Hull[2], David A. Cardwell[1] and John H. Durrell[1]

[1] Department of Engineering, University of Cambridge, Trumpington Street, Cambridge CB2 1PZ, UK
[2] The Boeing Company, Seattle, WA 98124-2207, USA

**Abstract**

A trapped magnetic field of greater than 3 T has been achieved in a single grain $GdBa_2Cu_3O_{7-\delta}$ (GdBaCuO) bulk superconductor of diameter 30 mm by employing pulsed field magnetisation (PFM). The magnet system is portable and operates at temperatures between 50 K and 60 K. Flux jump behaviour was observed consistently during magnetisation when the applied pulsed field, $B_a$, exceeded a critical value (e.g. 3.78 T at 60 K). A sharp $dB_a/dt$ is essential to this phenomenon. This flux jump behaviour enables the magnetic flux to penetrate fully to the centre of the bulk superconductor, resulting in full magnetization of the sample without requiring an applied field as large as that predicted by the Bean model. We show that this flux jump behaviour can occur over a wide range of fields and temperatures, and that it can be exploited in a practical quasi-permanent magnet system.

Bulk high-temperature superconductors (HTS) have been studied extensively in the pursuit of "permanent" magnet like materials with a magnetisation above that achievable using conventional rare earth compounds. As a result, Gruss *et al.* reported initially a trapped field of 16 T in an arrangement of two 25 mm zinc doped $YBa_2Cu_3O_{7-\delta}$ (YBCO) samples at 24 K [1], with a trapped field of 17.24 T later reported by Tomita *et al.* [2] between two YBCO samples. More recently, a field of 17.6 T has been achieved at the centre of a two sample stack of $GdBa_2Cu_3O_{7-\delta}$ (GdBCO) single grain bulk superconductors of diameter 25 mm [3]. The high energy density associated with these high trapped fields has aroused significant interest in potential applications for power systems, where a high magnetic field is required [4,5], and in superconducting levitation devices [6-8].

Significant technical challenges of bulk superconductors, however, remain in cooling and magnetisation technologies for practical applications. Pulsed field magnetization (PFM), which is a traditional method for activating permanent magnets, has been considered the most practical technique to magnetize HTS bulk superconductors either *in situ* or *ex situ* in portable systems [9,10]. However, the PFM technique is essentially a zero field cooling (ZFC) process, which implies that an external field of at least twice the peak field trappable in a bulk superconductor is required to achieve complete magnetization. Furthermore, since the magnetic field in PFM is applied for the order of milliseconds, significant practical challenges exist associated with the heat generation from the rapid, dynamic movement of magnetic flux in and out of the sample. Such effects can limit the efficiency of the PFM process and, consequently, the magnitude of the resulting trapped field [11-13].

A trapped field of 5.2 T in a bulk superconductor has been reported by Fujishiro *et al.* by employing a multi-pulse PFM technique with a maximum applied field of 6.7 T at an initial temperature of 30 K [14]. Recently, Weinstein *et al.*, very significantly, reported trapped fields close to 2 T at 77 K by applying pulsed fields comparable in magnitude to the trapped field [15,16]. We have also observed the sudden and rapid penetration of magnetic field into the centre of a bulk superconductor when the applied field reaches a critical value in a similar phenomenon to that reported by Weinstein and collaborators in their very high-$J_c$ samples. We consider the observed flux jump, or partial flux jump behaviour, [17-19] to be of great practical interest, since this provides potential for much easier PFM than was previously thought possible. The flux jump-assisted PFM process has been reproduced in numerical simulations using the finite element method (FEM), which suggests that a suitably large Lorentz force, $F_L$, can drive magnetic flux into the sample via a flux jump, which can result, in turn, in the sudden, full magnetization of the sample [20]. In this work, we report experimental investigations of this effect in a practical, portable system over a wide temperature range. In particular, we have achieved a trapped field greater than 3 T from a pulsed magnetic field of peak value $B_a$ = 4.86 T in a GdBCO single grain sample of diameter 30 mm at 51 K. In this case the Bean model predicts an external magnetic field of 17.8 T based on an average in-field $J_c$ of $7 \times 10^4$ A/cm$^2$ [22] for full penetration.

Figure 1(a) shows the portable magnet system. A 30 mm diameter, 12 mm thickness GdBCO bulk, single grain superconductor prepared via the conventional top seeded melt-growth (TSMG) technique [21] was installed in a copper sample holder located at the top of the vacuum can. Slots were made in the copper holder to reduce the formation of eddy currents during magnetization, as shown in figure 1(b). The bulk superconductor was cooled by conductive cooling from both its side and bottom surfaces. A portable Stirling Cryocooler (Cryotel CT, Sunpower) was used to provide the cooling power to a base temperature of 51 K. The bulk sample was characterised at 77 K by field cooled magnetization (FCM) to provide a reference. The peak trapped field was 1.01 T with a 2D magnetic field distribution, as shown in figure 1(c).

A Cernox sensor inserted into the copper sample mount was used to monitor the sample temperature. The applied and trapped fields were measured by a linear array of five Lakeshore HGT-2101 Hall sensors placed along one of the growth sector boundaries, at positions 0 mm, 2.5 mm, 5 mm 7.5 mm and 10 mm from the centre of the sample. The applied field, $B_a$, was generated via a copper solenoid coil placed outside the container of the portable system. The rise time of the pulsed field was 35 ms and the fall time was 500 ms for this system. Due to the fixed time duration, d$B_a$/d$t$ changes proportionally to $B_a$ with an initial value of 159 T/s for $B_a$ = 3.24 T and 171 T/s for $B_a$ = 3.51 T.

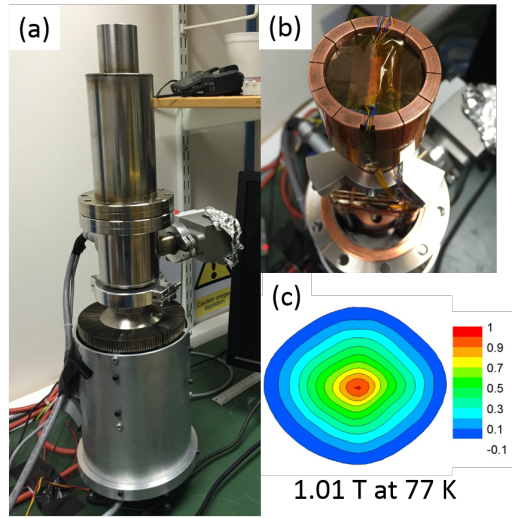

Figure 1. (a) Photograph of the portable magnet system, (b) a 30 mm diameter GdBCO bulk superconductor mounted in the slotted copper sample holder and (c) the 2D trapped field distribution mapped after FCM at 77 K.

The relationship between the trapped field after PFM and the applied field was investigated systematically over a wide temperature range. A critical applied field of 3.24 T at 70 K, 3.51 T at 65 K and 3.78 T at 60 K to fully magnetise the bulk superconductor was observed at each operating temperature, as shown in figure 2. Furthermore, it only required an increment of 0.27 T of the applied field to trigger the transition from partial to full magnetization of the sample. The inset to figure 2 reveals a more detailed dependence of the trapped field on the applied field at a specific operating temperature of 60 K. The applied field of $B_a$ = 3.78 T is assumed to be the optimal applied field, since further increments in $B_a$ only diminish the trapped field. This is consistent with previous experimental results reported by the present authors demonstrating a higher temperature rise due to a higher applied field [22-24].

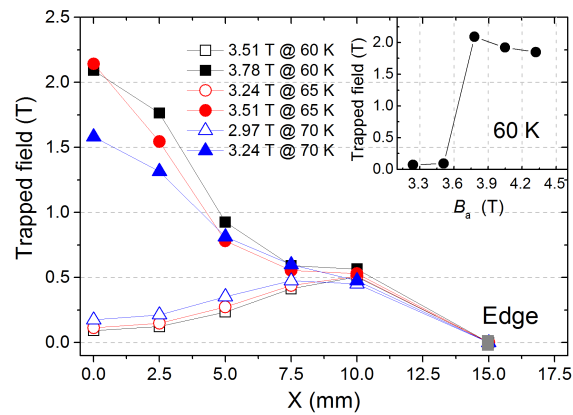

Figure 2. (a) Trapped field for the GdBCO bulk sample following the application of pulsed fields of different magnitude at temperatures of 60 K, 65 K and 70 K. A critical applied field to fully activate the bulk sample can be inferred from these data. The inset shows the peak trapped field, $B_T$, compared with the applied field, $B_a$, at 60 K. The data was taken 15 minutes after the PFM to minimise the influence from flux creep.

The variation of magnetic field during the entire PFM process was monitored with time during the magnetisation process to develop a better understanding of how flux lines penetrate into the sample. The shielding current at 65 K, and for $B_a$ = 3.24 T, prevented the flux from penetrating into the centre of the sample and resulted in the magnetic field being trapped near the edge of the sample, as shown in figure 3(a). The measured magnetic field at different positions on the surface of the sample was plotted during the rise and fall of the pulsed field during the PFM process. The resulting observed magnetic field distribution agrees well with the Bean model, with the gradient of the magnetic field representing the critical current density ($J_c$) of the sample, at least to a first approximation [25].

Increasing $B_a$ to 3.51 T resulted in a sudden rise in the field at the centre of the sample at a time of between 30 ms and 40 ms, which is assumed to originate from a flux jump. It then took more than one second for the flux to stabilise, with no significant fall in the trapped field measured during the flux creep period. It can be inferred from figure 3(d) that the flux jump occurs in the vicinity of the centre of the sample, and starts at a position at around 5 mm from the seed.

The "step" in magnetic field measured at 5 mm from the centre of the sample is observed at the same time as the sudden rise in magnetic field, as shown in the area highlighted by the dashed box in figure 3(b). This suggests that an instability first develops in this region and, consequently, the flux lines drift quickly through these areas and to the centre of the sample. A new equilibrium is then re-established with flux lines from the periphery of the sample moving subsequently into this region (i.e. that vacated by the flux lines displaced to the centre of the sample). This is similar to the flux jump behaviour observed in magnetization loops measured by SQUID magnetometry for single crystal [26] and melt-processed YBCO [27]. The sharp increase of the external magnetic field leads to severe heat generation, which heats up the sample in a localized area and consequently diminishes the $J_c$ within the framework of an adiabatic approximation [27,28]. As a result, the increase of the magnetic field at the diffusion front involves not only the increment of $B_a$ but also the reduction of the shielding field, which drives the instability. Mints *et al.* carried out a calculation on the critical value of applied field that would cause the flux jump, and found it to vary in proportion to the inverse of $(dB_a/dt)^{1/2}$ [29].

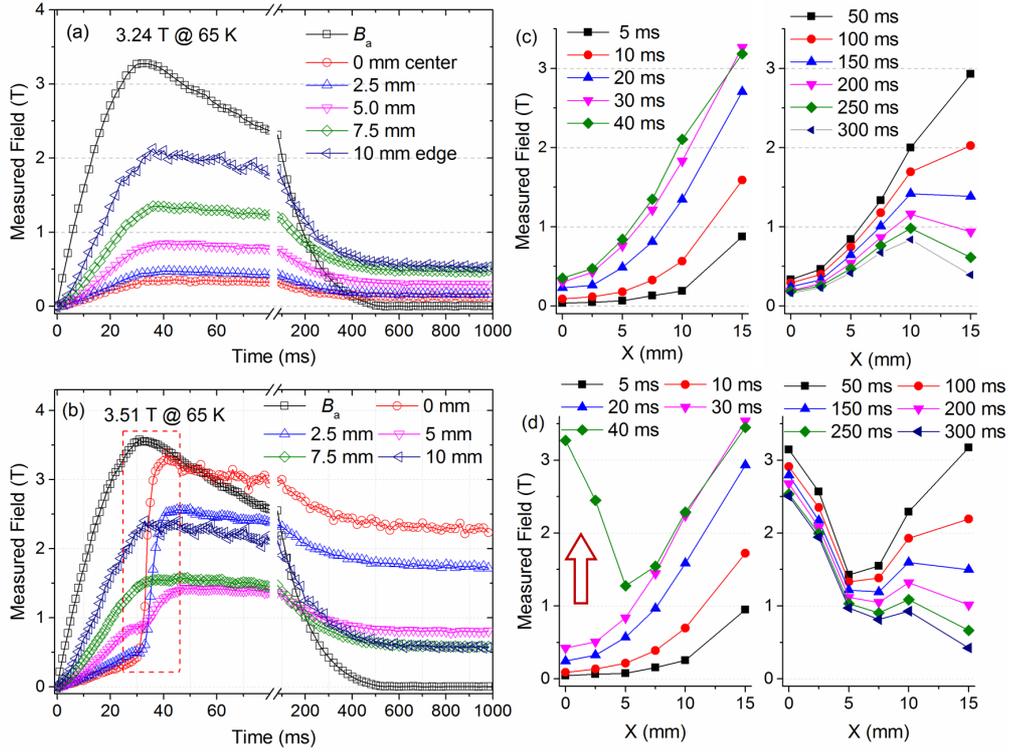

Figure 3. Magnetic field measured at the top-surface of the GdBCO bulk sample at 0 mm (centre), 2.5, 5, 7.5 and 10 mm (5 mm from the edge) during the PFM process at 65 K for $B_a$ = 3.24 T (a) and $B_a$ = 3.51 T (b). Magnetic field measured from the centre of the sample to the edge at different times during the rise and fall of the pulsed field for (c) $B_a$ = 3.24 T and (d) $B_a$ = 3.51 T. The slope, $dB/dx$, corresponds to $J_c$ according to the Bean model. The arrow indicates the position of the flux jump.

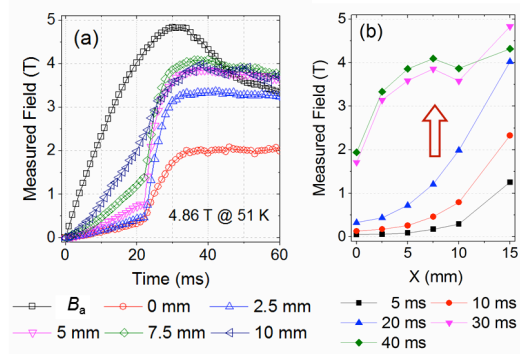

Figure 4. (a) Magnetic field measured at the top surface of the GdBCO bulk sample during the single pulse of maximum $B_a$ = 4.86 T at 51 K. (b) The variation of magnetic field across the surface of the sample during the rise of the pulse. The arrow indicates the position of the flux jump.

Figure 4 shows the magnetic field measured at the top surface of the GdBCO bulk sample during the rise of a single applied field of $B_a$ = 4.86 T at 51 K, and the variation of field across the surface of the sample during the rise of the pulse. Comparison of figures 3 and 4 indicates that the flux jump occurs earlier in the latter, at between $t$ = 20 and 30 ms, and the region where it occurs shifts towards the edge of the sample. A larger applied magnetic field, and consequently a larger d$B_a$/d$t$, generates more heat most of which is generated at the edge of the sample. This occurs since the thermal diffusion rate is much slower than that of the propagation of the electromagnetic fields during the PFM process [30,31]. This explains the occurrence of the flux jump at a position away from the centre of the sample.

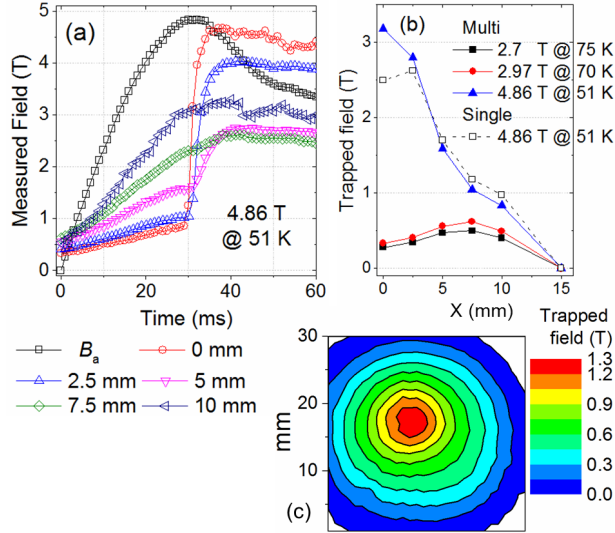

Figure 5. (a) Magnetic fields measured dynamically at the top surface of the GdBCO bulk sample during the rise of the final pulse at 51 K with $B_a$ of 4.86 T. (b) Comparison of the trapped field profiles achieved from single and multi-step (multi-pulse, multi-temperature) PFM processes (2.7 T @ 75 K → 2.97 T @ 70 K → 4.86 T @ 51 K). Partial magnetization of the sample is achieved at the pre-magnetization stage. (c) The distribution of the trapped field measured outside the vacuum can.

Pre-charging of magnetic flux half way to approximately the middle of the sample in the initial stage of a multi-step PFM process can effectively address the problem of heat generation near the edge of the sample during magnetization, thus facilitating the flux jump at the centre. This is consistent with previous research reported by Fujishiro *et al.* [12,32]. As shown in figure 5, in order to maximally pre-arrange magnetic flux towards the edge, a field of magnitude slightly lower than the critical field was applied to the sample, initially at 75 K, and then at 70 K. Finally, a field of 4.86 T was applied at 51 K. The magnetic flux was observed to jump all the way into the centre of the sample as a result of this process, generating a trapped field of 3.2 T. The trapped field was then mapped at the outer surface of the vacuum can at an absolute distance of 3.8 mm away from the sample surface. The distribution of the trapped field is rather uniform, as shown in figure 5(c).

The temperature of the magnetized sample was increased gradually to assess the stability of the trapped magnetic field. Flux creep started at 63 K, suggesting that 3.2 T is the maximum trapped magnetic field achievable in this sample at 63 K. This process is therefore equivalent to fully magnetising the bulk GdBCO superconductor at 63 K with a subsequent operating temperature of 51 K. This increses significantly the stability of the portable magnetic field as a result of stronger flux pinning at lower temperatures [33], which is particularly relevant for use in the presence of an applied external field [34,35].

In summary, we have observed that flux jump behaviour occurs consistently in the PFM process when the applied magnetic field exceeds a critical value, and that this critical value of field increases with decreasing temperature. The occurrence of the flux jump is found to be sensitive to the rate of application of the applied field, d$B_a$/d$t$. The flux jump effect is of significant technological importance since it enables magnetic flux to penetrate the bulk superconductor more easily than that predicted by the conventional Bean critical state model, as has been observed by Weinstein *et al.* We have demonstrated that this effect can be obtained in samples prepared by the conventional TSMG processing technique with much lower critical currents than in the samples used by Weinstein *et al.* and that the phenomenon can be exploited practically in order to achieve a magnetic field larger than 3 T for practical applications. We have built a portable magnet system that achieves such a field by exploiting this process.

This work was supported by the Boeing Company and by the Engineering and Physical Sciences Research Council (grant number: EP/P00962X/1). We are particularly gratefiul to Prof. Fujishiro of Iwate University for useful discussions. Data related to this publication are available at the University of Cambridge data repository http://dx.doi.org/10.17863/CAM.6290